\documentclass{article}

\usepackage{arxiv}

\usepackage[utf8]{inputenc} 
\usepackage[T1]{fontenc}    
\usepackage{hyperref}       
\usepackage{url}            
\usepackage{booktabs}       
\usepackage{amsfonts}       
\usepackage{nicefrac}       
\usepackage{microtype}      
\usepackage{lipsum}
\usepackage{graphicx}
\usepackage{amsmath}
\usepackage{biblatex}
\graphicspath{ {./images/} }

\title{Parameterizing Network Graph Heterogeneity using a Modified Weibull Distribution}

\author{
 Sinan A. Ozbay \\
  Bendheim Center for Finance\\
  Princeton University\\
  Princeton, NJ, USA\\
  \texttt{sozbay@princeton.edu} \\
   \And
 Maximilian M. Nguyen \\
  Lewis-Sigler Institute\\
  Princeton University\\
  Princeton, NJ, USA\\
  \texttt{mmnguyen@princeton.edu$^*$}
}

\begin{document}
\maketitle
\begin{abstract}
We present a simple method to quantitatively capture the heterogeneity in the degree distribution of a network graph using a single parameter $\sigma$. Using an exponential transformation of the shape parameter of the Weibull distribution, this control parameter allows the degree distribution to be easily interpolated between highly symmetric and highly heterogeneous distributions on the unit interval. This parameterization of heterogeneity also recovers several other canonical distributions as intermediate special cases, including the Gaussian, Rayleigh, and exponential distributions. We then outline a general graph generation algorithm to produce graphs with a desired amount of heterogeneity. The utility of this formulation of a heterogeneity parameter is demonstrated with examples relating to epidemiological modeling and spectral analysis.
\end{abstract}

\section*{Introduction}
Many real-world processes and phenomena can be accurately characterized as instances of graphs or dynamic processes on graphs \cite{newman_networks_2018}. Recent attention to graph models have been made in the areas of statistical models \cite{cheng_neural_1994, sarle_neural_1994, jordan_graphical_2004, goldenberg_survey_2010}, social network analysis \cite{scott_social_1988}, epidemiological models \cite{pastor-satorras_epidemic_2015}, political polarization models \cite{conover_political_2011}, and several other domains. For certain modeling problems, while it would be ideal to determine all of the nodes and edges of the graph empirically, the ability to do so is often limited in practice and prone to error. Thus, it is common to estimate the global or local topological features of the graph and then generate a graph matching those features on which simulations and computations can be performed and conclusions can be extrapolated.

One key feature of graphs that is of interest is the heterogeneity of a graph, which captures the amount of structural symmetry the graph encodes. To illustrate this intuitively, imagine a complete graph with N nodes, such that each node is connected to all other $N - 1$ nodes. Any process on this network would be indifferent to its particular starting point as each node is equally connected to all other nodes. Such a graph is considered homogeneous in its connectivity. On the other extreme, a graph where there is much more variation in what a single node is connected to encodes much more local structure. These types of graphs display a large amount of heterogeneity. Many processes that can be modelled with and simulated on graphs, such as epidemics, wild fires, diffusion problems, etc. are known to be sensitive to the amount of heterogeneity present in the graph modelling the system. As a result, having a rigorous and quantitative notion of heterogeneity that can be measured and controlled in simulations and calculations would be useful.

While qualitative comparisons between the heterogeneity of different types of graphs is common, it is difficult to make precise quantitative comparisons of the heterogeneity of graphs that generalize to many graph types for several reasons. First, there are many different types of graphs. Canonical examples include random graphs such as the Erdos-Renyi model, power-law graphs, small-world networks, complete and regular graphs, bipartite graphs, multiplex graphs, and many more that are of interest for varying real-world applications. Second, many of these graphs are generated using entirely different graph construction procedures. In light of the variation in how these graphs are generated, producing a parameterization that would allow for a simple interpolation between all of these different graphs is a seemingly difficult task.

A number of approaches to capturing heterogeneity are discussed in the literature. These include using the degree variance as a heterogeneity index \cite{snijders_degree_1981, bell_note_1992, smith_normalised_2020}, spectral approaches \cite{von_collatz_spektren_1957, estrada_quantifying_2010}, the Gini coefficient \cite{hu_unified_2008}, among others \cite{albertson_irregularity_1997, jacob_measure_2016}.

We note a particular approach that interpolates between two specific graphs, the Erdos-Renyi model and a scale-free network \cite{gomez-gardenes_scale-free_2006}. This method allows for a parameterization of heterogeneity wherein the first graph is very homogeneous in its degree distribution and the second is very heterogeneous. However, this interpolation method is limited in the two ways described above. First, this approach only allows for a comparison between two graphs at a time. Second, this method cannot necessarily be easily extended to consider other pairs of graphs since it interpolates at the level of graph construction, for which certain pairs of graphs would be strictly incompatible. However, as we will see below, by abstracting from the graphs themselves and instead focusing on the degree distributions, it becomes possible to quantitatively measure heterogeneity.

\section*{Method}

To solve the above problem, we present a control parameter that quantitatively represents the amount of heterogeneity in a graph as measured by the heterogeneity in its degree distribution. As the motivating example in the introduction shows, the heterogeneity is tightly, though not exactly, tied to the amount of local structure embedded in a graph, with the degree of nodes encoding all of the first order (or first neighbor) information about any given node. Thus, we make the key assumption that the heterogeneity of the graph can be accurately captured and quantified as a function of the degree distribution of the graph.

\begin{figure}
    \centering
    \includegraphics[width=0.48\textwidth]{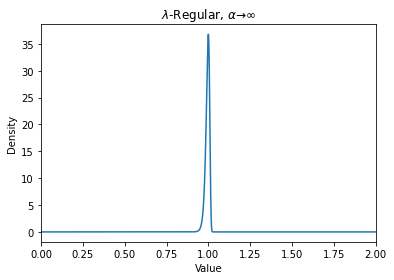}
    \includegraphics[width=0.48\textwidth]{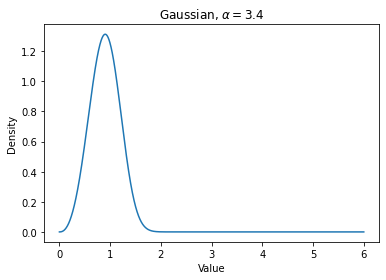}
    \includegraphics[width=0.48\textwidth]{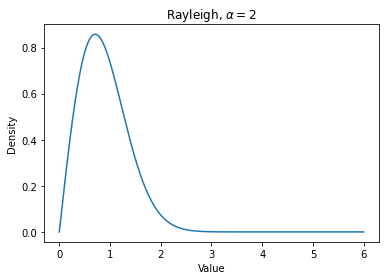}
    \includegraphics[width=0.48\textwidth]{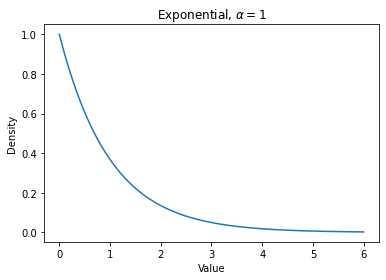}
    \includegraphics[width=0.48\textwidth]{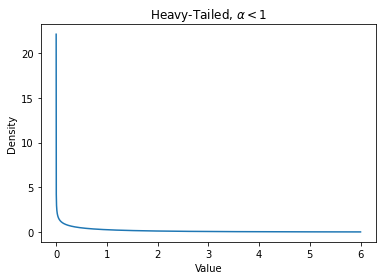}
    \caption{Various cases of the probability density functions of the Weibull distribution generated by varying shape parameter $\alpha$ and scale parameter $\lambda = 1$. a) $\alpha \to \infty$ produces a nearly singular distribution at $\lambda$. b) $\alpha=3.4$ produces a Gaussian-like distribution. c) $\alpha=2$ produces a Rayleigh distribution. d) $\alpha=1$ produces an exponential distribution. e) $\alpha < 1$ produces a heavy-tailed distribution.}
    \label{fig:cases}
\end{figure}

Once we make this assumption, we notice that the degree distributions of many graphs of empirical interest are given as special cases of the Weibull distribution. Recall the probability density function of the 2-parameter Weibull distribution \cite{rinne_weibull_2008}: 
\begin{align}
f(x; \lambda, \alpha) = \frac{\alpha}{\lambda}(\frac{x}{\lambda})^{\alpha - 1} e^{-(x/\lambda)^\alpha};   x\ge 0; \alpha, \lambda > \mathbb{R}^+   
\end{align}
where $\alpha$ is the shape parameter and $\lambda$ is the scale parameter. We note the following special cases of this function:

\begin{itemize}
\item $\alpha < 1$ corresponds to heavy-tailed degree distributions, with subexponential decay in the tail.
\item $\alpha = 1$ recovers the exponential distribution.
\item $\alpha = 2$ recovers the Rayleigh distribution.
\item $\alpha = 3.4$ corresponds to an approximately Gaussian distribution, which can recover the Erdos-Renyi graph.
\item $\alpha \to \infty$ recovers a $\lambda$-regular graph.
\end{itemize}
This can be seen graphically in Figure \ref{fig:cases}.

\begin{figure}
\centering
\includegraphics[scale=0.5]{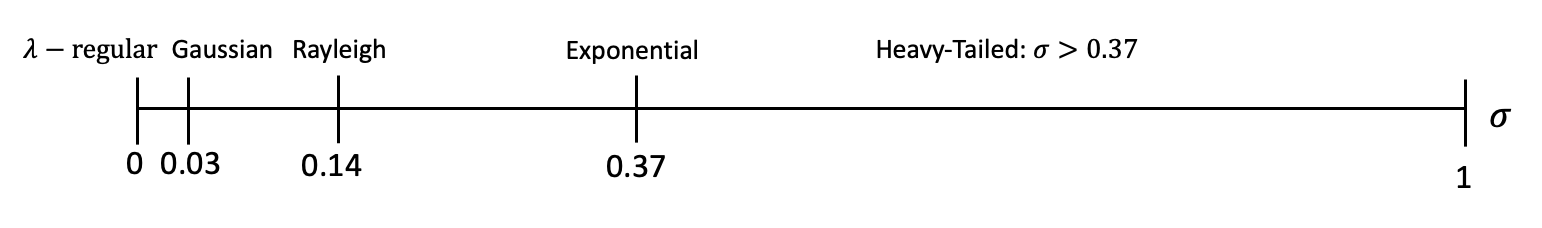}
\caption{Number line for heterogeneity that highlights the special cases corresponding to canonical distributions.}
\label{fig:line}
\end{figure}

We can further map the whole range of heterogeneity given by $\alpha$ to the unit interval using a simple transformation. We introduce a parameter $\sigma$. This parameter is given by:
\begin{align}
\sigma \equiv exp(-\alpha)  
\end{align}
where $\alpha$ is the shape parameter of the Weibull distribution. Since the shape parameter $\alpha$ is always a positive definite real number, under the exponential transformation this $\sigma$ parameter always lies in the range $(0,1)$. When $\sigma$ is close to $0$, the graph is homogeneous (more regular) in its degree distribution, and when $\sigma$ is close to $1$ the degree distribution is heterogeneous (more heavy-tailed). We can see the special cases in terms of a number line of $\sigma$ in Figure \ref{fig:line}.

The resulting exponential transformation of the shape parameter results in the following modified 2-parameter Weibull probability density function:
\begin{equation}
f(x; \lambda, \sigma) = \frac{-ln(\sigma)}{\lambda}(\frac{x}{\lambda})^{-ln(\sigma) - 1} e^{-(x/\lambda)^{-ln(\sigma)}};   x\ge 0; \sigma \in (0,1) ; \lambda > \mathbb{R}^+  \label{eqn:pdf}
\end{equation}

After we specify the desired level of heterogeneity in the graph, generating a degree distribution with that level of heterogeneity is simple. Two more parameters that do not relate to the heterogeneity of the graph must first be specified. These are $N$, the size of the graph in terms of number of nodes, and $\lambda$, the scale parameter, which can be thought of as the number at which the degree distribution will be centered at. These will vary from application to application. The values of these parameters will typically be estimated empirically. For instance, the size of the graph might be large if the modeller is considering a network the size of an entire country or small if instead a small neighborhood is being modelled. Similarly for the scale parameter, considering for instance social media networks, there might be data on the average number of friends users typically have.

Now with all of the parameters in hand, the following simple procedure generates a graph that has the desired heterogeneity: 
\begin{enumerate}
    \item Choose values of $\sigma$ (heterogeneity), $\lambda$ (center of degree distribution), and $N$ (number of nodes).
    \item Draw N random samples from the modified Weibull distribution specified in (\ref{eqn:pdf}) using $\sigma,\lambda, N$.
    \item Round each of the N samples to the nearest integer, since the degree of a node can only take on integer values.
    \item With the sampled degree distribution from the previous step, now use the configuration model method \cite{newman_networks_2018} (which samples from the space of all possible graphs corresponding to a particular degree distribution) to generate a corresponding graph.
\end{enumerate}  
This yields a valid graph $G(\sigma, \lambda, N)$ with the desired amount of heterogeneity as specified by $\sigma$.

\section*{A Numerical Example in Epidemiology}

We now present a simple numerical example from epidemiology which demonstrates the utility of the heterogeneity parameter $\sigma$. 

One of the key quantities or epidemic statistics of interest in the study of the spread of disease is the herd immunity threshold. This is the fraction of the population that has been infected by the time the number of infections peak. It measures the severity of the epidemic. It has been observed that the herd immunity threshold, in the context of epidemic simulations on graph networks, are quite sensitive to the amount of heterogeneity present in the contact network between people, which is represented by a graph.

Below, we present the results of SIR simulations of epidemics run on graphs of size N = 1000 and $\lambda = 5$, where the parameter of interest is $\sigma$. Each curve, from red to green, represents a different value of the graph heterogeneity. We implement the following simulation procedure:

Given a graph $G(\sigma,\lambda,N)$, fix a transmission probability $\tau$ and recovery probability $\gamma$:

\begin{enumerate}
\item At time $t_0$, fix a small fraction $f$ of nodes to be chosen uniformly on the graph and assign them to the Infected state. The remaining $(1 - f)$ fraction of nodes start as Susceptible. 
\item For each $i \in [1,T]$ where $T>>1$, for each pair of adjacent S and I nodes, the susceptible node becomes infected with probability $\tau$.
\item For each $i \in [1,T]$, each infected node recovers with probability $\gamma$.
\item At time $T$, record the herd immunity threshold at the peak of the epidemic.
\item Repeat steps (1-4) n = 150 times for each value of $\tau$.
\item Repeat steps (1-5) for each value of $\sigma$.
\end{enumerate}

\begin{figure}
    \centering
    \includegraphics[width=0.48\textwidth]{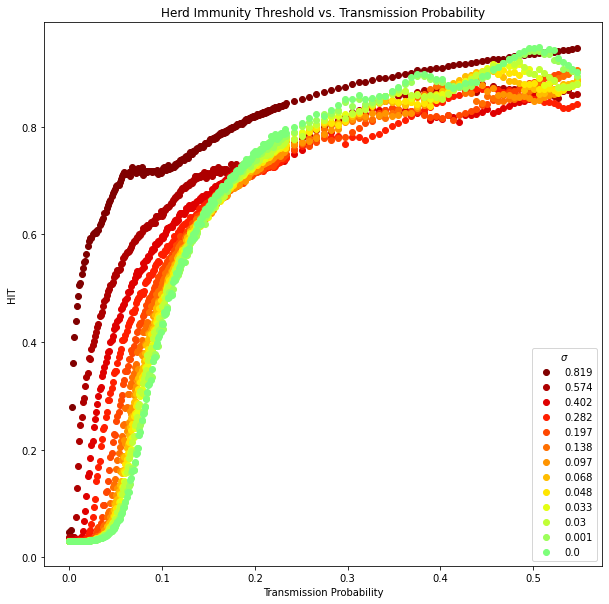}
    \caption{The herd immunity threshold for SIR epidemic simulations on networks with varying levels of heterogeneity ($\sigma$) as a function of transmission probability. $\lambda = 5, N = 1000$. Points represent the average of n = 150 simulation runs.}
    \label{fig:hit}
\end{figure}

The results are shown in Figure \ref{fig:hit}. This sensitivity analysis highlights two features that vary with heterogeneity. First, the red curves, with higher heterogeneity, lack a sub-critical regime that the more homogeneous curves have. This is recapitulates well-known behavior described in scale-free networks \cite{pastor-satorras_epidemic_2001}. Second, the HIT is not monotonic in the transmission probability for all graph types. Meanwhile, classical intuition suggests the proportion of graph reached at peak infections to be higher on average as the transmission probability increases. This counter-intuitive behavior is explored more by Ozbay et. al. \cite{ozbay_bifurcations_2022}.

This unexpected behavior was easily uncovered with a sensitivity analysis of this parameterization of heterogeneity. Because different graphs of interest are special cases corresponding to specific values of $\sigma$, we can easily interpolate between many different graphs and do so in a way that is precise and quantitative.

\section*{Analytical Calculations Using the Degree Distribution}

Using the probability density function given by (\ref{eqn:pdf}), the moments of the degree distribution can be expressed analytically. For example, the mean ($<k>$) and variance ($<k^2>$) of the degree distribution are as follows:
\begin{align}
<k> &= \lambda\cdot\Gamma(\frac{-1}{ln\sigma}+1) \label{eqn:mean}\\ 
<k^2> &= \lambda^2\cdot(\Gamma(\frac{-2}{ln\sigma}+1)-(\Gamma(\frac{-1}{ln\sigma}+1))^2) \label{eqn:var}
\end{align}
Thus depending on the particular problem, one could in principle make theoretical predictions of the impact of heterogeneity using the standard techniques of calculus. 

To illustrate this conceptually, we first note that there are closed form equations for a number of quantities associated with the configuration model expressed in terms of the degree distribution \cite{newman_networks_2018}. Examples of such calculations include the following: the clustering coefficient, the existence of the giant component, the mean size of a component of a randomly chosen node, the critical occupation probability for the giant percolation cluster, and the expected largest eigenvalue. 

\begin{figure}
    \centering
    \includegraphics[width=0.48\textwidth]{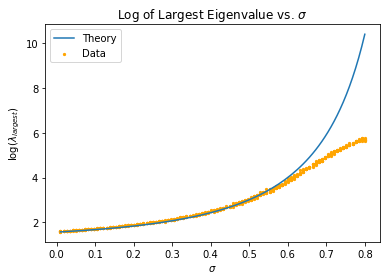}
    \includegraphics[width=0.48\textwidth]{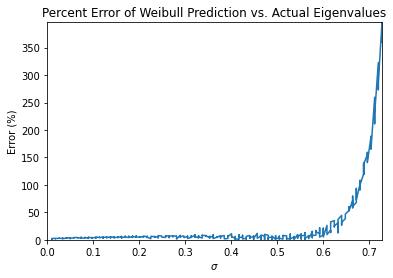}
    \caption{a) Comparison of the logarithms of the largest eigenvalue calculated by the theory of \cite{newman_networks_2018} and the empirically calculated value as a function of heterogeneity, $\sigma$ for system size N = 5000. b) Percent absolute error between predicted eigenvalue and empirically calculated eigenvalue, for various values of $\sigma$}.
    \label{fig:eig}
\end{figure}

As an example, we will characterize the sensitivity of the largest eigenvalue of a graph to changing graph heterogeneity. In the large network limit, the theoretical prediction for the largest eigenvalue in the configuration model ($\kappa_{1,theory}$) is given by the ratio of the variance to the mean of the degree distribution \cite{newman_networks_2018}. Using (\ref{eqn:mean}) and (\ref{eqn:var}) gives the following prediction.
\begin{equation}
\kappa_{1,theory}=\frac{<k^2>}{<k>}=\frac{\lambda^2\cdot(\Gamma(\frac{-2}{ln\sigma}+1)-(\Gamma(\frac{-1}{ln\sigma}+1))^2)}{\lambda\cdot\Gamma(\frac{-1}{ln\sigma}+1)} = \lambda [\frac{\Gamma(\frac{-2}{ln\sigma}+1)-(\Gamma(\frac{-1}{ln\sigma}+1))^2}{\Gamma(\frac{-1}{ln\sigma}+1)}] \label{eqn:eig}
\end{equation}
Figure \ref{fig:eig} shows both the theoretical prediction of the largest eigenvalue from (\ref{eqn:eig}) and the ground truth calculated empirically from generated graphs of the same parameter values $G(\sigma, \lambda, N)$. We see towards the homogeneous limit, the theory and data agree remarkably well. As heterogeneity increases, the discrepancy between theory and experiment grows quickly, but some amount of convergence still appears as system size grows (Figure S1).

\section*{Discussion}

Graphs are expressive modelling tools that can be, by the same token, complicated and difficult to characterize quantitatively in all of their features. In particular, one of the key advantages of graphs is that they can capture highly local information and interactions in a system. This makes the heterogeneity of the degree distribution a quantity of great interest for various applications of graphs and dynamical processes on graphs. The parameter $\sigma$ presented allows for several canonical graphs to be considered as special cases and easy interpolation between them. This improves on traditional methods of interpolation, where it was previously difficult to make direct comparisons. As the examples showed, processes on graphs and global statistics can be sensitive to local heterogeneity, necessitating a quantitative formulation of graph heterogeneity for those problems.

The key to the method proposed here is the choice of the Weibull distribution as a basis for deriving the different graphs. While other parameterizations using alternative statistical distributions are certainly possible, we justified the use of the Weibull distribution on the following grounds. A quantitative measure or parameterization of graph heterogeneity should ideally possess two qualities: expressiveness and interpretability. By expressiveness, we mean the parameterization should be broad enough to incorporate many of the cases of interest. There are many canonical graphs, so ideally a measure of heterogeneity should be able to assign a value to a reasonable number of them. One way to incorporate more possible graphs is to construct a more complex statistical distribution. There is a whole zoology of different statistical distributions. And with enough parameters one could construct a procedure that could generate arbitrary numbers of different graphs. However, this increasing complexity in parameterization comes at the cost of interpretability. Ideally it would be preferable to have a parameterization using a single scalar value as the effect of variation in that scenario is easy to interpret. One can make straightforward apples-to-apples comparisons between graphs in such a situation. In contrast, the interpretation of comparing two graphs when the parameterization is vector-valued is more complicated. Clearly there is a trade-off that occurs between expressiveness and interpretability, and a balance must be struck. In the context of a control parameter for heterogeneity, the modified Weibull distribution with its parameter $\sigma$ provides a balanced trade-off in terms of interpretability and expressiveness. As demonstrated above, many graphs of interest are captured strictly in terms of a single heterogeneity parameter.

\begin{figure}
    \centering
    \includegraphics[width=0.48\textwidth]{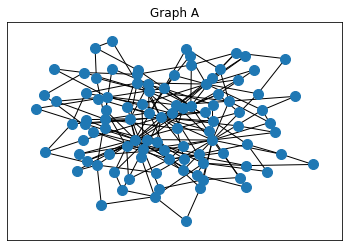}
    \includegraphics[width=0.48\textwidth]{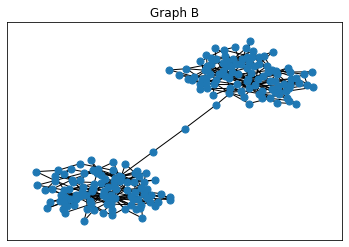}
    \includegraphics[width=0.48\textwidth]{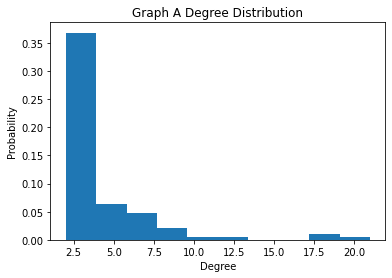}
    \includegraphics[width=0.48\textwidth]{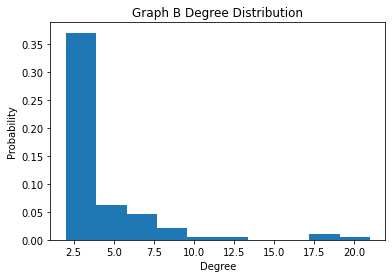}
    \caption{a) Visualization of a Albert-Barab\'asi graph with 100 nodes total and 2 nodes preferentially attached to each node. b) A barbell graph constructed by creating a copy of Graph A and attaching the two copies via a path graph. c) Histograms of Graph A degree distribution. d) Histogram of Graph B degree distribution.} 
    \label{fig:barbell}
\end{figure}

The main limitation of the stated parameterization of network heterogeneity is also its primary assumption, that the degree distribution of a graph will generally provide a complete picture of its overall topological heterogeneity. This assumption does not always hold true. In particular, the degree distribution only determines first-order and thus highly local topological properties about the graph. As can be seen in Figure \ref{fig:barbell}, we provide an example of two graphs with nearly identical degree distributions that are substantially different in their global topological features. In the example given, this manifests as Graph A lacking any protrusions or visible modularity, while Graph B has two distinct modules joined in a "barbell"-like shape. By focusing only on the degree distribution, our graph generation procedure is decoupled to the presence or absence of these large-scale features. Clearly though, such differences in global topological heterogeneity can have meaningful effects on stochastic processes on graphs. Despite their global structure being very different, both Graph A and B are in the space of graphs that can be generated by the configuration model, which only requires the degree distribution as input. While it is possible that Graph B could be constructed by the configuration model in principle, in practice Graph B has a very low probability of being constructed under this (or any random graph) construction procedure. That is because, for a given degree distribution, whether or not any two nodes are connected is completely random under the configuration model. Thus with a sufficiently large enough network, in the process of randomly connecting nodes it becomes extremely unlikely that roughly half of the nodes would not share any connections with the other half of the nodes. Therefore, graphs with some level of modularity make up a vanishingly small proportion of the space of graphs that are generated by this procedure. In contrast, real world networks can display substantial modularity which might be lost by only characterizing a degree distribution. Future work may look for a generalization of this approach beyond the degree distribution or configuration model.

In conclusion, the parameter $\sigma$ presented a simple means of controlling heterogeneity in a graph quantitatively which, in turn, can be used to better and more precisely understand the effect of changing graph topology on processes involving graphs.

\printbibliography

@article{goldenberg_survey_2010,
	title = {A {Survey} of {Statistical} {Network} {Models}},
	volume = {2},
	issn = {1935-8237, 1935-8245},
	url = {https://www.nowpublishers.com/article/Details/MAL-005},
	doi = {10.1561/2200000005},
	language = {English},
	number = {2},
	urldate = {2022-10-07},
	journal = {Foundations and Trends® in Machine Learning},
	author = {Goldenberg, Anna and Zheng, Alice X. and Fienberg, Stephen E. and Airoldi, Edoardo M.},
	month = feb,
	year = {2010},
	note = {Publisher: Now Publishers, Inc.},
	pages = {129--233},
}

@article{pastor-satorras_epidemic_2015,
	title = {Epidemic processes in complex networks},
	volume = {87},
	url = {https://link.aps.org/doi/10.1103/RevModPhys.87.925},
	doi = {10.1103/RevModPhys.87.925},
	number = {3},
	urldate = {2022-10-07},
	journal = {Reviews of Modern Physics},
	author = {Pastor-Satorras, Romualdo and Castellano, Claudio and Van Mieghem, Piet and Vespignani, Alessandro},
	month = aug,
	year = {2015},
	note = {Publisher: American Physical Society},
	pages = {925--979},
}

@book{newman_networks_2018,
	title = {Networks},
	isbn = {978-0-19-252749-3},
	language = {en},
	publisher = {Oxford University Press},
	author = {Newman, Mark},
	month = jul,
	year = {2018},
	note = {Google-Books-ID: YdZjDwAAQBAJ},
}

@misc{sarle_neural_1994,
	title = {Neural {Networks} and {Statistical} {Models}},
	author = {Sarle, Warren S.},
	year = {1994},
}

@article{cheng_neural_1994,
	title = {Neural {Networks}: {A} {Review} from a {Statistical} {Perspective}},
	volume = {9},
	issn = {0883-4237},
	shorttitle = {Neural {Networks}},
	url = {https://www.jstor.org/stable/2246275},
	number = {1},
	urldate = {2022-10-07},
	journal = {Statistical Science},
	author = {Cheng, Bing and Titterington, D. M.},
	year = {1994},
	note = {Publisher: Institute of Mathematical Statistics},
	pages = {2--30},
}

@article{jordan_graphical_2004,
	title = {Graphical {Models}},
	volume = {19},
	issn = {0883-4237, 2168-8745},
	url = {https://projecteuclid.org/journals/statistical-science/volume-19/issue-1/Graphical-Models/10.1214/088342304000000026.full},
	doi = {10.1214/088342304000000026},
	number = {1},
	urldate = {2022-10-07},
	journal = {Statistical Science},
	author = {Jordan, Michael I.},
	month = feb,
	year = {2004},
	note = {Publisher: Institute of Mathematical Statistics},
	pages = {140--155},
}

@article{snijders_degree_1981,
	title = {The degree variance: {An} index of graph heterogeneity},
	volume = {3},
	issn = {0378-8733},
	shorttitle = {The degree variance},
	url = {https://www.sciencedirect.com/science/article/pii/0378873381900149},
	doi = {10.1016/0378-8733(81)90014-9},
	language = {en},
	number = {3},
	urldate = {2022-10-07},
	journal = {Social Networks},
	author = {Snijders, Tom A. B},
	month = jan,
	year = {1981},
	pages = {163--174},
}

@article{bell_note_1992,
	title = {A note on the irregularity of graphs},
	volume = {161},
	issn = {0024-3795},
	url = {https://www.sciencedirect.com/science/article/pii/002437959290004T},
	doi = {10.1016/0024-3795(92)90004-T},
	language = {en},
	urldate = {2022-10-07},
	journal = {Linear Algebra and its Applications},
	author = {Bell, F. K.},
	month = jan,
	year = {1992},
	pages = {45--54},
}

@article{von_collatz_spektren_1957,
	title = {Spektren endlicher grafen},
	volume = {21},
	issn = {1865-8784},
	url = {https://doi.org/10.1007/BF02941924},
	doi = {10.1007/BF02941924},
	language = {de},
	number = {1},
	urldate = {2022-10-07},
	journal = {Abhandlungen aus dem Mathematischen Seminar der Universität Hamburg},
	author = {Von Collatz, Lothar and Sinogowitz, Ulrich},
	month = dec,
	year = {1957},
	pages = {63--77},
}

@article{jacob_measure_2016,
	title = {Measure for degree heterogeneity in complex networks and its application to recurrence network analysis},
	doi = {10.1098/rsos.160757},
	journal = {Royal Society Open Science},
	author = {Jacob, R. and Harikrishnan, K. P. and Misra, R. and Ambika, G.},
	year = {2016},
}

@article{scott_social_1988,
	title = {Social {Network} {Analysis}},
	volume = {22},
	url = {https://doi.org/10.1177/0038038588022001007},
	doi = {10.1177/0038038588022001007},
	number = {1},
	journal = {Sociology},
	author = {Scott, John},
	year = {1988},
	note = {\_eprint: https://doi.org/10.1177/0038038588022001007},
	pages = {109--127},
}

@article{albertson_irregularity_1997,
	title = {The {Irregularity} of a {Graph}},
	volume = {46},
	journal = {Ars Comb.},
	author = {Albertson, Michael O.},
	year = {1997},
}

@article{hu_unified_2008,
	title = {Unified index to quantifying heterogeneity of complex networks},
	volume = {387},
	issn = {0378-4371},
	url = {https://www.sciencedirect.com/science/article/pii/S0378437108001258},
	doi = {10.1016/j.physa.2008.01.113},
	language = {en},
	number = {14},
	urldate = {2022-10-07},
	journal = {Physica A: Statistical Mechanics and its Applications},
	author = {Hu, Hai-Bo and Wang, Xiao-Fan},
	month = jun,
	year = {2008},
	pages = {3769--3780},
}

@article{smith_normalised_2020,
	title = {Normalised degree variance},
	volume = {5},
	issn = {2364-8228},
	url = {https://doi.org/10.1007/s41109-020-00273-3},
	doi = {10.1007/s41109-020-00273-3},
	language = {en},
	number = {1},
	urldate = {2022-10-07},
	journal = {Applied Network Science},
	author = {Smith, Keith M. and Escudero, Javier},
	month = jun,
	year = {2020},
	pages = {32},
}

@article{estrada_quantifying_2010,
	title = {Quantifying network heterogeneity},
	volume = {82},
	url = {https://link.aps.org/doi/10.1103/PhysRevE.82.066102},
	doi = {10.1103/PhysRevE.82.066102},
	number = {6},
	urldate = {2022-10-07},
	journal = {Physical Review E},
	author = {Estrada, Ernesto},
	month = dec,
	year = {2010},
	note = {Publisher: American Physical Society},
	pages = {066102},
}

@article{conover_political_2011,
	title = {Political {Polarization} on {Twitter}},
	volume = {5},
	copyright = {Copyright (c) 2021 Proceedings of the International AAAI Conference on Web and Social Media},
	issn = {2334-0770},
	url = {https://ojs.aaai.org/index.php/ICWSM/article/view/14126},
	language = {en},
	number = {1},
	urldate = {2022-10-07},
	journal = {Proceedings of the International AAAI Conference on Web and Social Media},
	author = {Conover, Michael and Ratkiewicz, Jacob and Francisco, Matthew and Goncalves, Bruno and Menczer, Filippo and Flammini, Alessandro},
	year = {2011},
	note = {Number: 1},
	pages = {89--96},
}

@book{rinne_weibull_2008,
	address = {New York},
	title = {The {Weibull} {Distribution}: {A} {Handbook}},
	isbn = {978-0-429-14257-4},
	shorttitle = {The {Weibull} {Distribution}},
	publisher = {Chapman and Hall/CRC},
	author = {Rinne, Horst},
	month = nov,
	year = {2008},
	doi = {10.1201/9781420087444},
}

@article{pastor-satorras_epidemic_2001,
	title = {Epidemic {Spreading} in {Scale}-{Free} {Networks}},
	volume = {86},
	url = {https://link.aps.org/doi/10.1103/PhysRevLett.86.3200},
	doi = {10.1103/PhysRevLett.86.3200},
	number = {14},
	urldate = {2022-10-20},
	journal = {Physical Review Letters},
	author = {Pastor-Satorras, Romualdo and Vespignani, Alessandro},
	month = apr,
	year = {2001},
	note = {Publisher: American Physical Society},
	pages = {3200--3203},
}

@article{gomez-gardenes_scale-free_2006,
	title = {From scale-free to {Erdos}-{R}{\textbackslash}'enyi networks},
	volume = {73},
	url = {https://link.aps.org/doi/10.1103/PhysRevE.73.056124},
	doi = {10.1103/PhysRevE.73.056124},
	number = {5},
	urldate = {2022-11-28},
	journal = {Physical Review E},
	author = {Gómez-Gardeñes, Jesús and Moreno, Yamir},
	month = may,
	year = {2006},
	note = {Publisher: American Physical Society},
	pages = {056124},
}

@misc{ozbay_bifurcations_2022,
	title = {Bifurcations in the {Herd} {Immunity} {Threshold} for {Discrete}-{Time} {Models} of {Epidemic} {Spread}},
	url = {http://arxiv.org/abs/2212.06995},
	doi = {10.48550/arXiv.2212.06995},
	urldate = {2022-12-20},
	publisher = {arXiv},
	author = {Ozbay, Sinan A. and Nielsen, Bjarke F. and Nguyen, Maximilian M.},
	month = dec,
	year = {2022},
	note = {arXiv:2212.06995 [physics, q-bio]},
}

\section*{Acknowledgements}
The authors would like to acknowledge the members of the Levin Lab for their suggestions and feedback.

\section*{Author Information}
\subsection*{Authors and Affiliations}
\textbf{Bendheim Center for Finance, Princeton University, Princeton, USA} \newline
Sinan A. Ozbay

\textbf{Lewis-Sigler Institute, Princeton University, Princeton, USA} \newline
Maximilian M. Nguyen

\subsection*{Author Contributions}
S.A.O. and M.M.N. designed research, performed research, and wrote and reviewed the manuscript.

\subsection*{Corresponding author}
All correspondence should be directed to Maximilian Nguyen (mmnguyen@princeton.edu).

\section*{Availability of Data and Materials}
The code and datasets used and/or analysed during the current study are available from the corresponding author on reasonable request.

\section*{Ethics Declaration}
\subsection*{Competing Interests}
The authors declare no competing interests.

\end{document}


\begin{center}
\textbf{\large Supplemental Materials: Parameterizing Network Graph Heterogeneity using a Modified Weibull Distribution}
\end{center}

\renewcommand{\thefigure}{S\arabic{figure}}
\setcounter{figure}{0}

\begin{figure}[ht]
    \centering
    \includegraphics[width=0.33\textwidth]{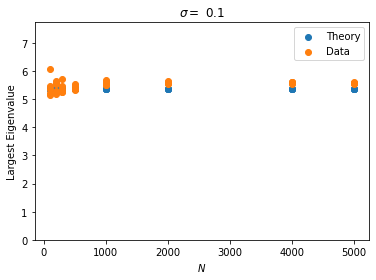}
    \includegraphics[width=0.33\textwidth]{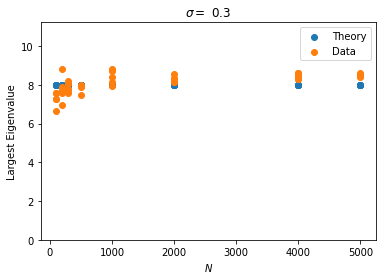}
    \includegraphics[width=0.33\textwidth]{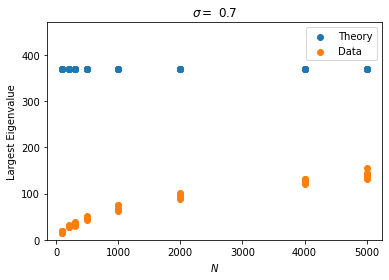}
    \caption{Comparison of the largest eigenvalue calculated by the theory of \cite{newman_networks_2018} and the empirically calculated value as a function of system size, $N$. Three levels of heterogeneity are shown, $\sigma = [0.1, 0.3, 0.7]$.} \label{fig:eigenvalueSize}
\end{figure}

\subsection*{Figure \ref{fig:eigenvalueSize}. Dependency of Largest Eigenvalue Prediction on System Size.}

\printbibliography